\documentclass[sigconf,nonacm,balance,screen]{acmart}
\usepackage[utf8]{inputenc}
\usepackage[T1]{fontenc}
\usepackage{balance}
\usepackage{graphicx}
\usepackage{color}
\usepackage{booktabs}
\usepackage{listings}
\usepackage{xspace}
\usepackage{array}
\usepackage{cuted}
\usepackage{capt-of}
\usepackage{enumitem}
\usepackage{microtype}
\usepackage{natbib}
\setcopyright{none}
\newcommand{\TODO}[1]{{\color{blue} TODO: #1}}

\begin{document}

\title{ThingTalk: An Extensible, Executable Representation Language for Task-Oriented Dialogues}
\author{Monica S. Lam \quad Giovanni Campagna \quad Mehrad Moradshahi \quad Sina J. Semnani \quad Silei Xu}
\affiliation{
    \institution{Computer Science Department}
    \institution{Stanford University}
    \city{Stanford}
    \state{California}
    \country{USA}
    \postcode{94305}
}
\email{{lam,gcampagn,mehrad,sinaj,silei}@cs.stanford.edu}
\renewcommand{\shortauthors}{Monica S. Lam, Giovanni Campagna, Mehrad Moradshahi, Sina J. Semnani, Silei Xu}

\begin{abstract}
Task-oriented conversational agents rely on semantic parsers to translate
natural language to formal representations. In this paper, we propose the design and rationale of the ThingTalk formal representation, and how the design improves the development of transactional task-oriented agents.

ThingTalk is built on four core principles: (1) representing user requests directly as executable statements, covering all the functionality of the agent, (2) representing dialogues formally and succinctly to support accurate contextual semantic parsing, (3) standardizing types and interfaces to maximize reuse between agents, and (4) allowing multiple, independently-developed agents to be composed in a single virtual assistant. ThingTalk is developed as part of the Genie Framework that allows developers to quickly build transactional agents given a database and APIs.

We compare ThingTalk to existing representations: SMCalFlow, SGD, TreeDST. Compared to the others, the ThingTalk design is both more general and more cost-effective. Evaluated on the MultiWOZ benchmark, using ThingTalk and associated tools yields a new state of the art accuracy of 79\% turn-by-turn.
\end{abstract}

\maketitle

\section{Introduction}
\label{sec:intro}

\begin{figure}
\centering
\includegraphics[width=0.8\linewidth]{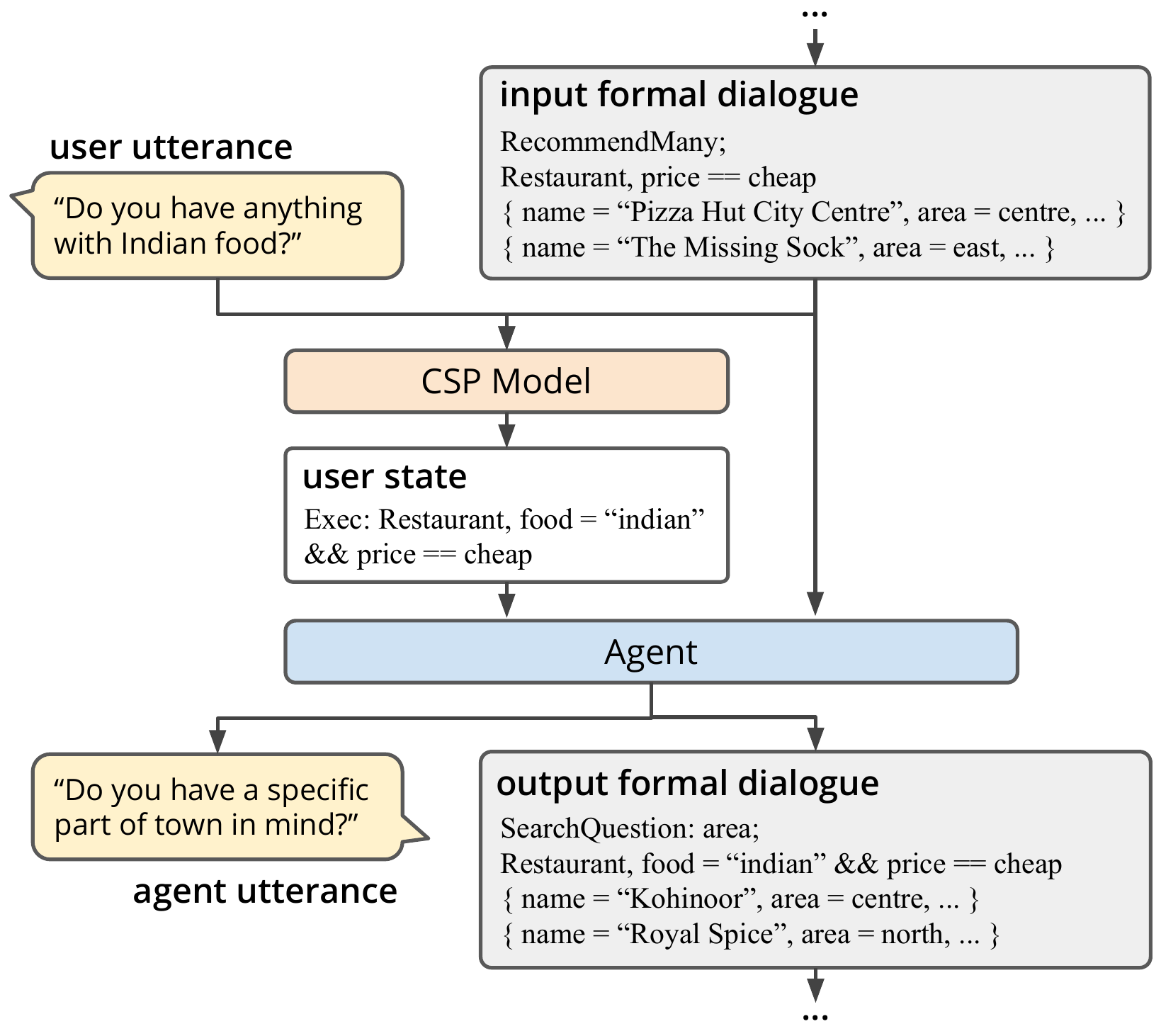}
\caption{Operation of a conversational agent using the ThingTalk representation. A contextual semantic parser is used to interpret the user input given a formal representation of the dialogue. The agent executes the representation and computes the reply.}
\label{fig:dialogue-loop}
\end{figure}

This paper presents the design and rationale of the ThingTalk dialogue representation language. ThingTalk is designed to support the synthesis of training data, effective semantic parsing, and the implementation of the agent.  All the tools developed with ThingTalk are grammar driven and can thus support extensions on ThingTalk. ThingTalk, the tools, the datasets, the trained neural semantic parsers, and the agents themselves are open-source and freely available.  We have used ThingTalk for many different tasks; this white paper focuses on the design and rationale of Thingtalk for transactional dialogues.  

\subsection{Purposes of ThingTalk}

ThingTalk is designed for these purposes. 
\begin{enumerate}
\item 
ThingTalk captures all the pertinent information of a dialogue needed by the agent to perform the tasks requested by the user. 
In fact, as an executable language, the agent can simply execute the program to achieve the effect. It is {\em complete} with respect to the agent's function.  The language must therefore be extensible in order to support the evolution of an agent's function. We have developed a ThingTalk {\em compiler and run-time system} to support this function.  

\item 
ThingTalk is designed as a target of the semantic parser from users' natural language utterances. It should match the mental model of a consumer, rather than that of a programmer.
In particular, we use a {\em contextual neural semantic model} for translating natural language into ThingTalk.  The model uses the formal ThingTalk dialogue representation as a context to interpret the current user's utterance. We have developed a GenieNLP library to scalably train and deploy the contextual semantic parsing model; the library is also applicable to other sequence to sequence neural models. 

\item 
ThingTalk is designed to support sample-efficient training, where most of the training data are to be synthesized.  We have developed a {\em Genie Toolkit} that leverages pretrained language models to automatically synthesize multilingual dialogue training data. 
\end{enumerate}

\subsection{Applications of ThingTalk}
We have used ThingTalk for many different tasks; they include: 
\begin{enumerate}
\item Task-oriented dialogues where users take the initiative to perform desired transactions, modeled as a composition of queries of knowledge bases and execution of a set of APIs ~\cite{dlgthingtalk}. 

\item
Virtual assistant tasks which may include monitoring streams of events and executing time-based or event-driven operations ~\cite{geniepldi19}. 

\item 
Web-based tasks which involve iterations or conditional executions of web operations, such as filling in text and pushing buttons ~\cite{10.1145/3453483.3454046}. 

\item
Sharing of data and commands among virtual assistants belonging to two or more users. This requires fine-grain access control and remote execution of ThingTalk programs~\cite{commaimwut18}. 
\end{enumerate}

\subsection{This Paper}
This paper focuses on the design and rationale of ThingTalk for transactional dialogues. ThingTalk is a more precise representation than previously proposed languages, covering 98\% of MultiWOZ. It also enables sample-efficient training with synthesis: using ThingTalk, we achieve a new state-of-the-art of 79\% on a reannotated MultiWOZ test set, using only 2\% of the normal amount of manually annotated training data.

In the rest of the paper, we first give a brief description of ThingTalk for representing transactional dialogues (Section~\ref{sec:overview}). We then discuss the rationale behind the language design (Section~\ref{sec:rationale}). We discuss the tools associated with ThingTalk and applications of ThingTalk beyond task-oriented dialogues in Sections~\ref{sec:tools} and \ref{sec:other}.
We evaluate by comparisons with other languages (Section~\ref{sec:comparison}) and on MultiWOZ (Section~\ref{sec:eval}), and conclude.  

\section{Overview of ThingTalk}
\label{sec:overview}

In this section, we provide a brief introduction to the ThingTalk language. A full description of the language is available in the ThingTalk documentation\footnote{\url{https://wiki.genie.stanford.edu/thingtalk}}.

\subsection{Turn Representation}
ThingTalk represents each turn in a dialogue using a \textit{dialogue act}, chosen from a set defined by the specific task-oriented agent, optionally followed by an executable representation of a request from the user or agent. Examples of ThingTalk representations are shown in Fig.~\ref{fig:user-state-example}.

\paragraph{Transaction dialogue acts} For transaction dialogues, the dialogue acts are chosen to model abstractly the behavior of the user and agent, independently of any specific domain.
We model a transaction dialogue as one where the user takes the initiative. As the user does not know what is available, this may involve querying one or more knowledge bases to decide on what interests them.
Hence, we identify the following kind of dialogue acts: 
\begin{enumerate}
\item The user and agent exchange pleasantries such as saying ``hello'' and ``thank you''. 
\item The user issues a query on the knowledge base, or refines the query based on the results the agent presented to the user.
\item The agent helps the user refine the query, by asking questions and offering results.
\item With back-and-forth conversation with the agent, the user conveys the parameters to the transaction of interest, and confirms or rejects the command.
\item The agent reports the result of executing the transaction to the user, either successful or failed.
\end{enumerate}

\paragraph{Representing requests}
In most turns of a transaction dialogue, the user issues a request that the agent executes a query or perform an action. ThingTalk represents the request as list of executable \textit{statements}. Each statement has this form:
\begin{align*}
    q \Rightarrow a ;
\end{align*}

$q$ is a \textit{query} to a database and $a$ is an API action with side-effect. Data flows from the query to the action.  Either the query or the action can be omitted. If the action is ommitted, the result is shown to the user.

Each of the query and action invokes a function defined in a class inside a \textit{skill library}. Each class has a globally unique identifier, using reverse DNS notation, such as ``@com.yelp'' or ``@com.twitter''. The class definition includes the parameters of each function and their type, chosen from a standard set of types or user defined. Classes can be abstract and can inherited to extend functionality. The class also declares how each function is implemented, i.e. whether it uses a standard protocol, or it requires custom code. Examples of class definitions are shown in Fig.~\ref{fig:class-example}.

The query supports common database operators: selection, projection, join, aggregation, sorting, ranking (Fig.~\ref{fig:user-state-example}). Selection filters can be specified using type-based operators on the fields of the database, combined with logical operators. Filters can also refer to other database tables, using a subquery. A filter can also specify that the user mentioned a parameter but they have no preference (``dontcare''), in which case the database query is not affected but the agent should not try to refine the parameter.

The action specifies parameters of the API by keywords. The value can specified as a variable, whose value refers to the output of the query, or directly as a constant. In the latter case, the constant is part of the user utterance, or copied from the context by the semantic parsing model.

\begin{figure}
\begin{tabbing}
1234\=123\=\kill
\textbf{Each command has a dialogue act:}\\
U:\>``thank you, that's enough for now.''\\
R:\>@Transaction.Cancel;\\
\\
\textbf{The Execute dialogue act requests an action from the agent:}\\
U:\>``find me an action movie.''\\
R:\>@Transaction.Execute;\\
R:\>$@\text{TMDB}.\text{Movie}(), \text{contains}(\textit{genres}, \text{``action''});$\\
\\
U:\>``post a tweet saying hello.''\\
R:\>@Transaction.Execute;\\
R:\>$@\text{Twitter}.\text{Post}(\textit{status}=\text{``hello''});$\\
\\
\textbf{Complex queries are possible:}\\
U:\>``find me the top 3 closest Chinese restaurants.''\\
R:\>@Transaction.Execute;\\
R:\>$\text{sort}(\text{distance}(\textit{geo}, \text{here})~~\text{asc}~~\text{of}~~@\text{Yelp}.\text{Restaurant}(),$\\
\>\>$\text{contains}(\textit{cuisines}, \text{``Chinese''}))[1:3];$\\
\\
\textbf{Queries and actions can be combined in one sentence:}\\
U:\>``play songs by Adele.''\\
R:\>@Transaction.Execute;\\
R:\>$@\text{Spotify}.\text{Song}(), \text{contains}(\textit{artists}, \text{``adele''})$\\
\>\>$\Rightarrow @\text{Spotify}.\text{Play}(\textit{song}=\textit{id})$
\end{tabbing}
\caption{Examples of ThingTalk representation (R) for a user command (U). Each representation starts with the dialogue act, followed by executable query-action statements. In the fifth example, the \textit{id} field of the Song table is passed to the action by name. The @ sign denotes a namespace for a dialogue act, query or action; namespaces are simplified for exposition. The examples use the @Transaction namespace for dialogue acts, as they are targeted at the transactional (voice assistant) dialogue agent.}
\label{fig:user-state-example}
\end{figure}

\begin{figure}
\begin{tabbing}
123\=12345678901\=\kill
\textbf{abstract} \textbf{class} @MediaPlayer \{\\
\>\textbf{entity} Song, Artist, Genre;\\
\>\textbf{query} Song(\textbf{out} \textit{id} : Entity(Song),\\
\>\>\textbf{out} \textit{artists} : Array(Entity(Artist)),\\
\>\>\textbf{out} \textit{genres} : Array(Entity(Genre)),\\
\>\>\textbf{out} \textit{release\_date} : Date,\\
\>\>\textbf{out} \textit{popularity} : Number);\\
\\
\>\textbf{query} CurrentlyPlaying(\textbf{out} \textit{id} : Entity(Song));\\
\\
\>\textbf{action} Play(\textbf{in} \textit{song} : Entity(Song));\\
\}\\
\\
\textbf{class} @Spotify \textbf{extends} @MediaPlayer \{\\
\>\textbf{loader} @Thingpedia.JavaScriptV2();\\
\>\textbf{query}~\text{Favorites}(\textbf{out} \textit{id} : Entity(Song));\\
\}
\end{tabbing}
\caption{Example of classes in the skill library. Each class include user-defined entity types, as well as declarations of queries and actions with their input and output parameters. In this example, the @MediaPlayer class declares a generic database of songs, a query for the currently playing song, and an action to play a song. The (concrete) @Spotify class extends the abstract @MediaPlayer class with Spotify-specific functionality. The @Spotify class also declares how it is implemented, in this case using the Thingpedia~\cite{almondwww17} format and JavaScript code.}
\label{fig:class-example}
\end{figure}

\subsection{Summarizing Dialogues}
In addition to representing individual turns of a dialogue, ThingTalk supports a succinct representation of a whole dialogue. The representation includes all the information necessary for the agent to the continue the dialogue, and for a contextual semantic parsing model to interpret the next turn. 

For a transaction dialogue, all the information is in the requests that the user issued and the agent executed. Hence, the dialogue can be summarized by the ThingTalk statements executed, along with the results, or partial statements that the user may be interested in. 

\subsection{Semantic Parsing of ThingTalk}
ThingTalk is designed to be used in an agent built using a contextual semantic parsing approach (Fig.~\ref{fig:dialogue-loop}). The formal representation of the dialogue is sufficient to interpret the user utterance: the dialogue representation, concatenated to the user utterance, is fed to a neural semantic parsing model to compute the formal representation of the turn. The agent appends the turn representation to the dialogue representation, executes it if possible, and then computes the reply to the user. All state is encoded in the formal dialogue representation and the model is free to predict any syntactically-valid representation, which the agent can react to. Hence, this design allows the user to change their mind or follow unexpected paths, making it more robust than traditional dialogue trees.

For training, ThingTalk is designed to support synthesis: a large space of dialogues is automatically generated, by having a simulator of the user interact with an agent that makes simulated API calls~\cite{dlgthingtalk}. The synthesized data is automatically paraphrased~\cite{xu2020autoqa} and used to fine-tune a pretrained BART sequence-to-sequence model~\cite{lewis2019bart}. The model is then further fine-tuned on a small set of manually annotated data (less than 50 times the commonly used amount of data), allowing the model to generalize beyond the state machine~\cite{dlgthingtalk}.

Both the agent and the user simulator are developed using a domain-independent state machine~\cite{campagna-etal-2020-zero}. The state machine is defined once and can be instantiated for many domains by providing the class definition. Each turn is synthesized by combining domain-independent templates associated with each control construct and operator in the language, with domain-specific phrases for each query, action, and parameter~\cite{geniepldi19,xu2020schema2qa}. A large library of templates is provided as part of the ThingTalk tools, covering database queries, transactional dialogues, as well as trigger-action programming.

Overall, the design of ThingTalk greatly reduces the development cost of new conversational agents: synthesis reduces the need for expensive crowdsourcing of data and manual annotation, and agent designers can reuse types and classes from an existing library to further reduce the engineering cost of dialogues.

\section{Design Principles of ThingTalk}
\label{sec:rationale}

\subsection{Standardized Types}
Our goal is to create virtual assistants that can handle many domains, each of which can be implemented by different vendors using different database and API schemas. Our approach is to establish a standard type system, borrowing as much as possible from open standard schemas such as Wikidata and Schema.org.

To support this goal, we designed ThingTalk as a strongly-typed language with many built-in types. ThingTalk supports the standard types (strings, booleans, numbers, arrays, ...). Second, ThingTalk natively supports  common basic data types: locations, dates, times, measurements. 
For every type, it includes the associated operators, such as ``distance'' for locations.
Third, ThingTalk has adopted standard entity types. For example, languages and countries are represented as ISO codes; companies are represented by their stock ticker.


The static type system allows to enforce semantic constraints on ThingTalk representations. The training data synthesizer only produces representations that type checks. The use of high-level, fine-grained types renders all valid type combinations meaningful, even if uncommon. Note that the semantic parser learns to predict properly typed representations directly from data. 

ThingTalk also encourages standardization at the level of whole agents. Using inheritance, specific skills can reuse and extend functionality from abstract agents in common domains. The abstract agents can be built collaboratively. We adopted the types used in Wikidata and Schema.org, two well-known schemas, to design abstract agents for question-answering.

%

Having standardized data types and schemas reduces the repetitive code and data between different conversational agents. Furthermore, it is important to support composition of functions independently developed, especially in different domains.  It also allows annotations and tools be shared across agents in different domains. Standardizing on these common schemas also supports transfer learning across different domains. 

\subsection{ThingTalk Execution Statements}
ThingTalk is designed for implementing transactional, task-oriented dialogue agents. Our goal is to cover everything that an agent can possibly do, therefore we need a programming language as an interface. It is inadequate to just recognize the slots mentioned by the user.

ThingTalk uses a small number of construct constructs to cover the general functionality of the agent. 
This factors out domain-dependent information from the domain dependent constructs. It shares the advantages of database design where all the tools (compilers, query optimizations, etc) developed for the small set of well-defined query primitives can be applied to all domains. 

\subsubsection{Integration of Database Queries and API}
Whereas many previous proposals focus mainly on queries or APIs,
ThingTalk has a unified syntax that integrates queries with API invocations, compositions, conditional executions, and streams. 
For queries, ThingTalk adopts relational semantics, and includes all the major relational operators: projection, selection, join (cross product), subquery (semijoin), aggregation. In addition, ThingTalk supports invocations of APIs and compositionality through parameter passing. Conditional execution is data-oriented, represented as filters, which are similar in notation for selection and projection in database queries (Fig.~\ref{fig:user-state-example}).
ThingTalk also includes operators for constructing streams of data by monitoring the result of a query.

To illustrate why unification of queries and APIs is important, consider the simple example of playing a named song by a specific artist. One would not expect a specific API accepting that particular combination of parameters. In ThingTalk, this is succinctly represented as a composition of a song query with two filters and a {\em play} API call.

\subsubsection{Canonicalization}
ThingTalk is \textit{canonicalizable}.  Syntactically different natural language utterances have the same ThingTalk representation if they have the same meaning. Canonicalization improves the accuracy of training, and it is also necessary for paraphrasing. Query constructs are canonicalized so operators are applied in a specific order (filters, sorting, ranking, aggregation, projections). Filters are normalized to conjunctive normal form and then sorted alphabetically. Keyword parameters are sorted.

\subsubsection{Extensibility}
To support new functionality in agents,
the constructs are implemented in a grammar-based manner in open-source tools. Developers can easily extend the language to new constructs, by providing the definition, runtime semantics, and a few natural language templates.

\subsection{State-Based Dialogue Representation}
\label{sec:dialogue-summary}
In ThingTalk, the whole dialogue is not represented as a history of utterances but a summary of the {\em state} of the conversation necessary to interpret the new user utterances. It includes the completed transactions with their results as well as ongoing transactions. Abandoned queries are not represented. To keep it simple for semantic parsing, we only include the last completed transaction and one outstanding transaction for each domain. The neural semantic parser resolves the coreferences and updates the state as the user changes the request, which we believe is more robust than rule-based systems. Furthermore, the semantic parser needs to only parse the last user utterance turn in natural language, with the compact dialogue state fed in as context. This improves accuracy in practice, because it reduces the amount of training data needed. A short input also allows using standard pretrained neural models, which have a fixed input length, further improving accuracy.

\subsection{Cross-Vendor Function Composition with A Global Namespace}
ThingTalk allows agent developers to create agents that compose multiple skills, potentially developed by third-party. This makes it possible to support virtual assistants with hundreds or thousands of available commands - many more than the five to ten domains of typical task-oriented systems. Combining skills is possible because each skill in ThingTalk is loaded from a library and namespaced with a globally unique namespace based on DNS. Furthermore, all skills use the same ThingTalk types, allowing data to be passed between them. 

Compared to a commercial virtual assistant, ThingTalk is unique because it combines all skills in a single model, rather than relying on an out-of-band dispatch system. Once the skills are combined in one agent, multi-domain interactions become possible, such as using the result of a query in one skill as the parameter to an action in a different skill.

The use of a global namespace also enables discovering skills dynamically using web technology, rather than entering them in a centralized repository ahead of time. This allows ThingTalk to operate in a decentralized manner: different developers can provide conversational agents to be used in the same conversation, sharing data between each other.

\section{Tools}
\label{sec:tools}
 ThingTalk includes extensive tooling in the form of the Genie Toolkit for conversational agents~\cite{geniepldi19}. The tools are fully open-source and available to any developer to extend for their own use case.
 
\subsection{Tooling for Annotation}
ThingTalk tools include static checks for annotation conventions in manually annotated data. Without clear annotation conventions, and the ability to enforce them automatically, different dialogues will be annotated inconsistently, which will negatively affect accuracy. Experience with MultiWOZ evidences the need for conventions on the normalization of values and types in the ontology, and how mentioned slots are carried over from one turn to the next. ThingTalk has a well-defined syntax of each type, and statically enforces rules on how values connect to utterances. For example, numbers and strings in a formal representation must appear identical either in the context or in the utterance at the same turn.
 
\subsection{Natural Language Templates}
As ThingTalk is defined by a grammar, we can use grammar rules or templates to either translate ThingTalk into a natural language, or to synthesize equivalent ThingTalk and natural language expressions. After synthesis, the data can be paraphrased, either manually by crowdworkers or automatically~\cite{xu2020autoqa}.

To provide additional diversity in synthesis time, we have created about 900 {\em domain-independent templates} that map ThingTalk types, operators and control constructs into a wide variety of natural language sentences.  By using ThingTalk's rich type system, templates can match the operator $\ge$ for ``temperature'' to ``hotter than'', and for ``distance'' to ``farther than''. Having such diversity reduces the need of paraphrasing, which can introduce both errors and cost, especially if manual paraphrasing is used.  



\subsection{Abstract State Machines}
The strong emphasis of ThingTalk in separating domain-independent constructs from domain knowledge makes it possible for us to define an abstract domain-independent state machine for modeling dialogues. Using the state machine, we can automatically synthesize training data for a transactional agent, given database schemas and API signatures.

Note that synthesized dialogues from state machines are only a subset of what can be represented by ThingTalk. Thus, while dialogues are synthesized from state machines, agents are expected to handle the full generality of ThingTalk. Our results on MultiWOZ show that this is very important: while ThingTalk can represent 98\% of MultiWOZ, the state machine can only represent 85\%~\cite{dlgthingtalk}. For the remaining 15\%, the agent must rely on the generalization capabilities of the model, enhanced with a small amount of manually annotated data.

\section{Other Applications of ThingTalk}
\label{sec:other}

As mentioned in Section~\ref{sec:intro}, ThingTalk has been designed with more than task-oriented dialogues in mind. Here we briefly highlight how the design was applicable in two representative use cases.

\paragraph{Event-driven commands}
ThingTalk includes a \textit{monitor} construct that allows to trigger when the result of any query changes, supporting event-driven use cases in a natural and compositional way~\cite{almondwww17,geniepldi19}. 

The use of a language construct makes monitoring available to all conversational agents, without requiring developers to reimplement a notification system every time.

\paragraph{Multi-agent communication}
A distributed ThingTalk protocol allows agents to communicate by issuing ThingTalk commands to each other, and receiving back results~\cite{commaimwut18}. For security, the user can specify access control rules in natural language, which are translated to a \textit{ThingTalk Access Control Language} formally establishing what commands are allowed to be executed and from whom.

Multi-agent communication extends the capability of a conversational agent beyond what a single organization can provide in a centralized system, further improving the user experience.
\section{Representation Comparisons}
\label{sec:comparison}

\subsection{SMCalFlow}

The Semantic Machines Dataflow Format (SMCalFlow)~\cite{SMDataflow2020} is another recently proposed dialogue representation. Like ThingTalk, it also represents the user intent as an executable representation. The representation is predicted from a contextual semantic parser and updates an executable dialogue context. The agent directly interprets the dialogue context to produce the answer to the user. 

Like any other programming language, ThingTalk programs implicitly define a dataflow graph, which can easily be generated with a compiler. Syntactically, however, we believe that it is easier to read programs than to read data flow graphs.

In this section, we highlight the major differences between  ThingTalk and SMCalFlow.

\subsubsection{Dialogue History (SMCalFlow) vs Dialogue State (ThingTalk)}
While ThingTalk summarizes the dialogue by its state (Section~\ref{sec:dialogue-summary}),
SMCalFlow is designed to capture the history of the dialogue formally, a turn at a time. 
Semantic parsers for SMCalFlow use a concatenation of all user utterances and all agent programs in the dialogue history as input to the semantic parser~\cite{SMDataflow2020}. The history grows with the length of the dialogue and cannot be easily compressed; this reduces generalization across turns and increases the amount of dialogue data necessary to learn the representation. Furthermore, encoding the entire history is incompatible with using modern pre-trained neural models that have a fixed input length.

In ThingTalk, the semantic parser needs to only parse the last user utterance turn in natural language, with the compact dialogue state fed in as context.

\subsubsection{Surface Semantics (SMCalFlow) vs State-Based Semantics (ThingTalk)}

SMCalFlow represents users' updates to the dialogue context at the surface level, in terms of explicit meta-computation operators ``refer'' and ``revise''. ThingTalk design is similar to the ``inlined'' design discussed by the SMCalFlow authors~\cite{SMDataflow2020}, which is to represent utterances with the resulting modified states. 

The ``revise'' operator is problematic for a dialogue representation language because edit programs are not unique: multiple syntactically distinct edit programs can yield the same dialogue context. 
A canonical formal representation of each turn is helpful to train sequence-to-sequence semantic parsers~\cite{geniepldi19}. 

ThingTalk's design is also more general, because it is not limited to particular edit operations. ThingTalk can naturally support multiple edit operations in the context; in SMCalFlow, this requires nesting multiple ``revise'' operation. 

Furthermore, unlike SMCalFlow, ThingTalk does not need any heuristic to perform the edit operation on the context at runtime. In ThingTalk, coreferences are represented by typed tokens (e.g. ``date\_0'', ``restaurant\_0'') that the model can learn to copy, while in SMCalFlow they are represented as ``refer'' node that must be resolved heuristically. ThingTalk learns the coreference implicitly from data. Heuristics can be encoded through synthesized data, while cases that deviate from the heuristic are supported through few-shot annotated data. The dialogue representation in ThingTalk is also normalized to make copying efficient. 


Empirically, experiments on SMCalFlow did not find any measurable difference between implicit and explicit representation of the edit operations on the MultiWOZ dataset~\cite{SMDataflow2020}. On the SMCalFlow dataset, they found the inlined representation yields a only 5\% lower accuracy, despite the dataset being constructed explicitly for SMCalFlow and not being optimized for copying.

\subsubsection{Completeness of Queries}
ThingTalk semantics are grounded in relational algebra (for the query part) and direct API calls (for the action part), regardless of the application domain. This factoring of constructs and domains supports modularity and generalization as discussed in Section~\ref{sec:rationale}. 

On the other hand, SMCalFlow does not have a construct with equivalent power to a database query. Instead, it only includes a simple ``constraint'' system based on pattern matching: a constraint is a record whose fields can be unspecified, specified with a value, or specified with a comparison operator.
This constraint system is weaker than a full query language; e.g. it does not support arbitrary logic (and, or, not). It also requires domain-specific logic for sorting, ranking, aggregation, etc..

\subsubsection{Annotation Complexity}

Both ThingTalk and SMCalFlow are hard to annotate. SMCalFlow is additionally more challenging to paraphrase, because of its representation of coreferences through the ``refer'' operator. Phrases with equivalent intent, and that would yield the same query by the agent, are annotated differently depending on whether they refer to entities implicitly or by name. Hence, one cannot paraphrase between the two sentence forms, because that would change the representation. This is difficult to ensure, either with crowdsourced or automatic paraphrasing.

\subsubsection{Data Synthesis}
Because ThingTalk is designed to support synthesis of training data, every ThingTalk construct comes with natural language templates, and it is possible to construct natural language sentences for arbitrary ThingTalk programs. SMCalFlow does not have a canonical natural language form~\cite{DBLP:journals/corr/abs-2104-08768}, 
nor does it have a library of templates for synthesis. Indeed, it cannot have a domain-independent template library, because it does not have any domain-independent semantics, types, or operators.

\subsubsection{Scalability}
The SMCalFlow dataset only includes four domains: calendar, weather, places, and people. ThingTalk instead has been validated on the five MultiWOZ domains~\cite{dlgthingtalk} and on 44 virtual assistant domains~\cite{geniepldi19}. ThingTalk can scale to many domains thanks to the comprehensive builtin type system and the library of constructs, greatly reducing the domain-specific effort needed to build a new domains. 

\subsubsection{Composability of Domains}
ThingTalk is designed to support virtual assistants: conversational agents with thousands or tens of thousands of skills, many of which are provided by third parties.
ThingTalk skills are namespaced and declared in a standard format, while SMCalFlow APIs are entirely ad-hoc. Hence, virtual assistant developers can easily compose ThingTalk skills, without fear of incompatibilities. They cannot easily do so in SMCalFlow. Furthermore, because ThingTalk offers a standard representation for common types, different skills can interoperate without changes.

\subsubsection{Open-Source}
Finally, ThingTalk is entirely open-source: both the language and the associated tools are developed by an open community. SMCalFlow is a proprietary language. Even the implementation for the SMCalFlow research dataset was not fully released, because it included proprietary code~\cite{SMDataflow2020}. The community needs an open-source language that they can extend, and needs open-source tools. A proprietary language cannot grow to meet everyone's needs.

\subsection{SGD}

Schema Guided Dialogues~\cite{rastogi2019towards} is a dialogue representation language previously proposed by Google.
SGD is based on a fixed set of dialogue acts, combined with API calls from a schema.

SGD is similar to ThingTalk because:
\begin{itemize}
\item It enforces well-defined conventions for representing types, including exact, canonical representation of categorical (enumerated) types.
\item It includes a standard schema representation that allows different skills to be combined interoperably.
\end{itemize}

SGD differs from ThingTalk because:
\begin{itemize}
\item It only supports APIs, not database queries. Hence, compositional requests need different APIs. The SGD representation is not compositional and not well-suited to questions.
\item While SGD supports template-based synthesis, it only includes a very limited template system, designed to be used with human paraphrasing. Synthesized data cannot be used to train directly. Human paraphrasing is expensive, and it is desirable to avoid it.
\item SGD was only evaluated on paraphrased data obtained from a state machine designed in concert with the representation itself. It is not clear that the representation generalizes to WOZ data, let alone real data.
\item SGD includes only limited open-source tooling as part of the dataset release. 
\end{itemize}

\subsection{TreeDST}
Previous work by Apple proposed TreeDST, a compositional representation for dialogues~\cite{cheng-etal-2020-conversational}. Like ThingTalk, TreeDST captures a full dialogue formally and is suitable for use in a semantic parser.

TreeDST differs from ThingTalk because:
\begin{itemize}
\item TreeDST is not directly executable. Agents need domain-specific logic to interpret the TreeDST representation. This increases the engineering cost to build a new agent.
\item Like SGD, TreeDST was only evaluated on paraphrased data obtained from a state machine designed in concert with the representation itself. The representation does not generalize to WOZ data, let alone real data. In particular, the choice of dialogue act is too limited because the simulated data never includes a turn without a request, such as a plain ``hello''.
\item TreeDST includes no tools with the dataset release, nor it is documented beyond the aforementioned paper.
\end{itemize}

\subsection{Others}
The Amazon Alexa assistant is known to use the Alexa Meaning Representation Language (AMRL)~\cite{kollar2018alexa}. AMRL is a graph representation where nodes are grounded in some token of the dialogue utterances. Different sentences with identical semantics have different representations; hence, AMRL needs to be manual annotated and cannot be synthesized and paraphrased. Furtheremore, AMRL has no defined executable semantics; each skill interprets the representation in a domain-specific fashion. 

Similar to AMRL, the TOP dataset from Facebook proposed a lexically grounded, compositional semantic representation~\cite{DBLP:journals/corr/abs-1810-07942}. TOP is a tree representation where every node is a span of the utterance; hence, as in AMRL, syntactically different sentences have different representations. TOP is designed to be manually annotated by crowdworkers, but it is not suitable for synthesis. TOP also does not model dialogue, it only supports representing single utterances.

Previous work also proposed using the AMR language for dialogues~\cite{bai2021semantic, bonial2020dialogue}. AMR is a general meaning representation applicable to any text. It is unsuitable for a task-oriented dialogue because its notion of semantics is (1) too fine-grained and tied to the underlying natural language syntax to be practical, and (2) inadequate for an agent to act on, because it models the surface meaning of the sentence and not the underlying user intention.
\section{Experimental Evaluation}
\label{sec:eval}
ThingTalk for task-oriented dialogues was evaluated on the MultiWOZ benchmark~\cite{dlgthingtalk}. We reannotated the 2.1 test set, and part of the validation set, constructing MultiWOZ version 3.0 where every turn is annotated with ThingTalk (user, agent, results). Annotation was performed by experts, and is significantly more accurate than previous annotation efforts.

ThingTalk could precisely model 98\% of the test set. The remaining 2\% were mainly human errors by the WOZ crowdworkers, and out-of-domain utterances.

A contextual semantic parser (CSP) trained on ThingTalk achieved 79\% turn-by-turn exact match accuracy, in a few-shot setting that used only 2\% of the normal amount of training data. When comparing only slots, the CSP model achieved 87\% accuracy, well above the state of the art of 61\% on MultiWOZ 2.1. This highlights the advantage of a precise and correct representation.

On the validation set, the training strategy with synthesis and automatic paraphrasing improves over training with only manual annotated data by 6\% exact match, and 7\% on slots. This highlights the advantage of synthesis, and the need to design the representation to support synthesis.

\section{Conclusion}
\label{sec:conclusion}

In this white paper, we have presented ThingTalk, a novel formal representation for dialogues that enables creation of more expressive conversational agents with lower cost.

ThingTalk has been shown effective in MultiWOZ, leading to a new state of the art accuracy of 79\% on the reannotated test set. It was also shown effective for event-driven commands and access control specifications in natural language.

ThingTalk is part of Genie, a large, collaborative effort to build a open-source, privacy-preserving virtual assistant. Genie includes both a usable virtual assistant supporting the 10 most popular skills, as well as tools for developers to build new skills and new agents scalably, and deploy them on a platform of their choice. ThingTalk\footnote{\url{https://github.com/stanford-oval/thingtalk}} and Genie\footnote{\url{https://github.com/stanford-oval/genie-toolkit}} are open-source and available on GitHub.

\begin{acks}
We would like to thank all the students and open-source contributors who helped developed the ThingTalk language and tools over the years.

This work is supported in part by the National Science Foundation under Grant No. 1900638, the Alfred P. Sloan Foundation under Grant No. G-2020-13938, and the Verdant Foundation. 
\end{acks}

\bibliography{main}


\end{document}